\documentclass{article} 
\usepackage[utf8]{inputenc}
\usepackage{graphicx}
\usepackage[table,xcdraw]{xcolor}
\setkeys{Gin}{width=\columnwidth}
\usepackage[subpreambles=true]{standalone}
\usepackage{appendix}
\usepackage{tikz}
\usepackage{circuitikz}
\usepackage{float}
\usepackage[backend=biber, url = false, firstinits]{biblatex}
\usepackage{algorithm}
\usepackage{algpseudocode}

\title{High tension lines: Measuring power-grid robustness under variable line loads using graph representation}

\author{Jonathan Bourne}
\date{February 2021}

\begin{document}

\maketitle

\begin{abstract}

This paper explores whether graph embedding methods can be used as a tool for analysing the robustness of power-grids to cascading failures within the framework of network science. The paper focuses on the strain elevation tension spring embedding (SETSe) algorithm and compares it to node2vec, Deep Graph Infomax, and the measures mean edge capacity and line load. These five methods are tested on how well they can predict the collapse point of the giant component of a network under random attack. Seven power-grid networks are used, ranging from 14 to 2000 nodes. A total of 3456 loading profiles are generated on each power-grid, simulating a wide range of conditions. One hundred random attack sequences are generated for each load profile, and the ability of each method to model the mean point of collapse is then compared using a generalised additive model. Only SETSe and line load perform well as proxies for robustness with both measures having and $R^2 = 0.89$ averaged across the seven networks. The attack process is repeated using 8784 load profiles representing typical yearly operation on the 2000 node network; line load performs exceptionally well ($R=0.99$), outperforming SETSe's best score ($R=0.80$). Finally, it is shown that due to its method of local smoothing and global weighting of node features, SETSe creates an interpretable geographical embedding that provides valuable qualitative insight into the state of the power-grid. 
Using just under 3.3 million attack to failure simulations, this paper shows that graph node representation algorithms can be used to analyse network properties such as robustness to cascading failure attacks.

\end{abstract}

\section*{Key words}
\begin{itemize}
    \item Networks
    \item Infrastructure networks
    \item graph embedding
    \item robustness
\end{itemize}

\section{Introduction}

power-grids are some of the most important network structures in modern society; without them, almost all areas of daily life would be affected. One of the greatest issues facing power-grids and critical infrastructure networks in general is that of cascading failures \cite{gray_super-blockers_2018, motter_cascade-based_2002, watts_simple_2002, ash_optimizing_2007}. In Ref.~ \cite{baldick_initial_2008} cascading failures are defined as ``a sequence of dependent failures of individual components that successively weakens the power system", these failures include physical and cyber-infrastructure as well as human error and organisational or procedural issues. For simplicity within the field of network science cascading failures are usually limited to failures in nodes and edges. As such, a simplified definition of a cascading failure is when an edge connecting two nodes in a network is removed either by failure/attack or through exceeding its maximum carrying capacity. The flow over the lost edge is then diverted, which can cause secondary failures. Any additional edges that fail are then removed from the network. This process repeats until no edges exceed their capacity and the cascade terminates. It is clear that the amount of power a network can handle at any one time -- its carrying capacity -- along with its topology, will affect the robustness of the network and the size of any cascade. However, understanding how to reduce cascades can be challenging \cite{dobson_complex_2004, newman_exploring_2011}. 

Although real power-grids use alternating current, researchers investigating cascading failures in power-grids often use DC power to simplify the power-flow calculations \cite{cuadra_critical_2015}. In this paper, we use the DC flow equations described in Refs.~\cite{pepyne_topology_2007, arianos_power_2009}. The calculation of DC power flow is significantly more straightforward than the calculation of AC power flow, as it requires only linear equations rather than an iterative process. The DC flow equations have the form $\mathbf{f}=\mathbf{CA(A^TCA)^{-1}p}$, where $\mathbf{A}$ is the adjacency matrix of the network with the slack bus removed to make the system invertible, $\mathbf{C}$ is a diagonal matrix of the line susceptance, and $\mathbf{p}$ is the power generated or demanded by each node. The slack bus absorbs or supplies additional power as required to ensure that the system is balanced.

Line limits are required to calculate cascades on the power-grid. However, due to a lack of real data, synthetic line limits are often generated using proportional loading \cite{motter_cascade-based_2002, kinney_modeling_2005, pepyne_topology_2007, zhu_revealing_2014, koc_impact_2014}. Proportional loading assumes the line limit is proportional to the flow over the edge at the start of the simulation. As such, the maximum flow over line $i$ is given by $f_{i,j}^\mathrm{max} = \alpha \left | f_{i,j}^c \right |$, where $f_{i,j}^c$ is the power flow over the line connecting nodes $i$ and $j$, and under initial flow conditions, $\alpha$ is the system tolerance. The system tolerance is a real-valued number in the range $1 \geq \alpha \geq \infty$. 

Understanding the robustness of power networks is a rich area of research \cite{kelly-gorham_using_2019, motter_cascade-based_2002, koc_impact_2014, kinney_modeling_2005}. Robustness itself is not an observable property but a latent one that is inferred using some metric and often using Monte-Carlo simulation. For a recent literature review on this topic, see Ref.~\cite{cuadra_critical_2015}. Many of the studies examining power-grid robustness have focused on networks that are proportionally loaded. In a previous study \cite{bourne_dont_2019}, we found that a poorly parametrised $\alpha$ can result in cascade behaviour that is very different from real line limits, and that such parametrisation is common in the literature.

An alternative approach to using specially designed robustness measures, would be to map the network itself to an alternate space, such that network properties are preserved, and use this embedded representation to reveal the robustness of the network. Embedding methods for tabular data are commonly used for dimensionality reduction \cite{pearson_liii_1901} and visualisation \cite{van_der_maaten_visualizing_2008}. However, embedding methods for networks are becoming more popular as a way of gaining insight into network structures and behaviour. Because network embedders can create a representation of the entire topology of the network, in some cases also including node features, it may be that this representation also includes information on latent graph properties such as network robustness, which otherwise can only be inferred from a large number of simulations. Some network embedders return a single measure for an entire network \cite{narayanan_graph2vec_2017, gutierrez-gomez_unsupervised_2019}, however, such methods cannot provide information related to nodes and edges, losing valuable information. This paper only considers methods that return node level embeddings. The paper builds on previous work which introduced the Strain Elevation Tension Spring embedding (SETSe) algorithm \cite{bourne_spring_2020-1}, a deterministic physics based model that represents node features as forces and network edges as springs. SETSe produces three different types of embedding output, node elevation from an initial plane, and edge embeddings which are measured in terms of mechanical strain and tension. 

The goal of the paper is to see whether the embedding methods are able to predict the collapse point of the giant component. The hypothesis is that networks which, when embedded by SETSe, have a high mean tension/strain will have a lower attack tolerance than networks with low tension/strain. To test how effective SETSe at this task, we will compare it against mean edge tolerance $\alpha$, mean line load where $\textrm{LL} = \frac{1}{\alpha}$ and two alternative network embedding approaches node2vec \cite{grover_node2vec_2016} and Deep Graph Infomax (DGI) \cite{velickovic_deep_2018}.

This paper performs three main analysis. The first analysis uses synthetic line limits to explore how well different approaches can represent the robustness of the power-grid across a wide range of load profiles. The second analysis uses time series data to explore how well robustness can be represented within `normal' grid operating parameters. The final analysis compares SETSe and line load as a qualitative method for analysing power-grids.

\section{Method}

This section begins by explaining the calculation of the robustness measures, a simple motivating example is provided showing the conceptual link between SETSe and power-grid robustness. This will be followed by an introduction to the datasets used. The method for creating the different line loads across the networks is then discussed, followed by the power-grids attack method. The evaluation criteria of how the 5 measures will be assessed is introduced. Next a description of the time series analysis will be given, finally the method geospatial analysis will be described.

\subsection{The different methods used as robustness measures}
\label{sec:diff_robust_measures}
As mentioned in the introduction, a standard metric used in research into cascading failures in the power-grid is the system tolerance $\alpha$, although its inverse the line load ($\mathrm{LL} = \frac{1}{\alpha}$) is also used. The system tolerance $\alpha$ is the mean edge capacity of the system, and as such it is susceptible to skew and can be made arbitrarily large, e.g., $\infty = \frac{1+ \infty}{2}$. The line load is similar to the harmonic mean of $\alpha$; its range is between 0 and 1 and it is a Schur-convex metric that is dominated by its smallest value and therefore cannot be made arbitrarily large $\frac{1}{2} = \frac{\frac{1}{1}+ \frac{1}{\infty}}{2}$. 

In this paper, the mean edge capacity and line load are compared with the embeddings produced by SETSe, node2vec \cite{grover_node2vec_2016}, and DGI \cite{velickovic_deep_2018}. node2vec and DGI produce node level embeddings, whilst SETSe produces node and edge embeddings, all three embedding methods take weighted networks, but only DGI and SETSe can make use of node features, which in this case is net power produced/demanded. Node2vec uses a random walk to embed each node in a vector space, the algorithm uses the transition probability distribution to maximise the likelihood of the local neighbourhood being preserved in the embedding space using stochastic gradient descent. Node2vec is valuable as it allow non-linear relationships between nodes. In contrast, DGI is based on the graph convolutional network developed in Ref.~\cite{kipf_semi-supervised_2016} and has been extended to be more appropriate for unsupervised embedding. DGI uses mutual information to avoid over emphasising the local neighbourhood, an issue that can affect node2vec. In contrast to the other two embedding methods SETSe cannot be said to `learn` and does not have a loss function in the conventional sense. Instead, SETSe finds a physical equilibrium across the network, where the embedded elevation values of the features are a function of all other node features with importance decreasing as a function of distance and magnitude and modified by the weighted topological structure of the network itself. The physical equilibrium results in node embeddings that are locally smoothed and globally weighted through the network topology. This provides a pleasing network example of Tobler's first law of geography that states `everything is related to everything else, but near things are more related than distant things' \cite{tobler_computer_1970}. 

A clear drawback of SETSe is that, whilst node2vec and DGI can embed graphs in an arbitrary number of dimensions, making them suitable for complex network topologies, SETSe embeds the graph in a fixed number of dimensions dependent on the number of node features. Despite this, SETSe has been shown to outperform node2vec and DGI in low dimensions \cite{bourne_spring_2020}, but similarly if DGI and node2vec are allowed a larger number of embedding dimensions. In this paper, node2vec and DGI will embed the power networks into a 32-dimensional vector space as opposed to the 1-dimensional space SETSe will use. 

SETSe is parametrised using the net normalised power of the node as the ``force" variable, such that $F_i = \frac{2G_i}{\sqrt{G^2_i} }$, where $G_i$ is the power generated or demanded by node $i$. The value $\alpha$ of each edge is transformed to a spring stiffness $k$ using $k_{i,j} = k_{\mathrm{range}}(1-\frac{1}{\alpha_{i,j}}) + k_{\mathrm{min}}$ where both the spring stiffness range $k_{\mathrm{range}}$ and the minimum spring stiffness $k_{\mathrm{min}}$ are positive real values. This paper will use arbitrarily chosen values of $k_{\mathrm{min}} = 100$ and $k_{\mathrm{range}} = 1000$ as this parametrisation has no impact on the results, see Appendix.~A. For a detailed description of the SETSe algorithm see Ref.~\cite{bourne_spring_2020-1}.

Both the node2vec and DGI implementations are taken from the StellarGraph library \cite{data61_stellargraph_2018}, whilst an R implementation of the SETSe algorithm is available from \url{https://github.com/JonnoB/rSETSe}.

\subsubsection{Converting node/edge values to global values}
As the methods used in this paper create values for individual edges and nodes,as such they need to be aggregated to create a value representing the entire network.
The proxies for robustness that will be used in this paper are the scalers mean edge tolerance ($\alpha$) and mean edge line load. From SETSe the robustness measures will be the scalers mean node absolute elevation, mean edge strain and mean edge tension. From the other embedding methods, node2vec and DGI, will use the vector of length 32 of mean node absolute value and also the scaler mean edge euclidean distance between all node pairs.
The scaler values will be calculated using $\bar{m} = \frac{\sum \left| m_i \right|}{n}$, where $m$ is the robustness measure, e.g., strain, $\alpha$, etc., and $n$ is the number of nodes or edges (depending on the metric). The euclidean distances will be calculated using $\bar{m} =\sum \frac{\left \| m_i-m_j \right \|}{n}$, where $m_i$ and $m_j$ are the vector embeddings of the connected nodes and $n$ is the total number of edges in the network.

For ease of comparison, all robustness measures will be bounded between 0 and 1 using $\kappa = \frac{\bar{m} - \bar{m}_{\mathrm{min}}}{\bar{m}_{\mathrm{max}}-\bar{m}_{\mathrm{min}}}$, where $\bar{m}_{\mathrm{min}}$ is the smallest mean value for each network and each robustness measure across all simulations (e.g., the smallest strain value produced on IEEE-14), and the value $\bar{m}_{\mathrm{max}}$ is the largest value produced for each robustness measure and network across all simulations. 

Having introduced the robustness measures the next section provides the intuition linking SETSe embeddings and the robustness of power-grids.

\subsection{Motivation: The link between system tolerance and strain}
\label{sect:motivation}

\begin{figure}[h!]
  \begin{center}
  \includegraphics{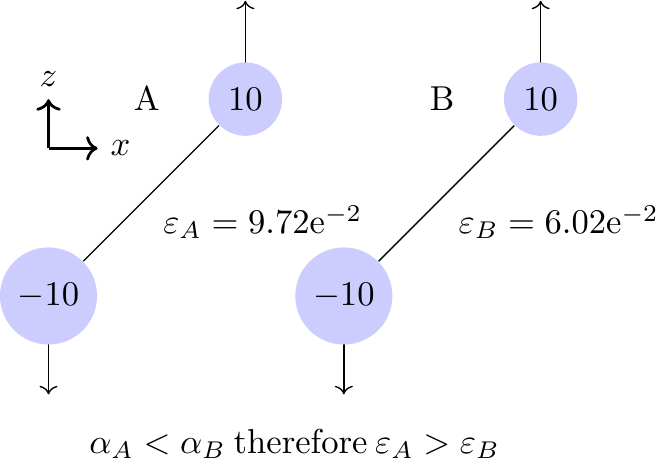}
    \caption{When networks A and B experience the same force but $\alpha_{\mathrm{A}}=1.2$ and $\alpha_{\mathrm{B}}=1.8$, it is intuitive that when embedded using SETSe the more highly loaded network experiences greater strain. The networks are shown in the $x$--$z$ plane.}
    \label{tikz:toy_example_1}
  \end{center}
\end{figure}

Consider the networks shown in Fig.~\ref{tikz:toy_example_1}. Both networks A and B have the same parametrisation of $k$ with $k_{\mathrm{min}}= 100$, $k_{\mathrm{range}} = 1000$, and an $x$--$y$ distance between the nodes of 1. However, network A has $\alpha = 1.2$; in contrast, network B has $\alpha = 1.8$; as such, A is more heavily loaded than B. These different loadings result in $k$ values of 250 and 500, respectively, and this means that, per unit force, A will stretch more than B. The vertical forces generated by the nodes are balanced in the two networks when A has a strain of $\varepsilon = 9.72\times 10^{-2}$ and B has a strain of $\varepsilon = 6.02\times 10^{-2}$. This can be proved simply using trigonometry.

It is clear that, for any two-node networks, $\varepsilon_{\mathrm{A}} > \varepsilon_{\mathrm{B}}$ when $\alpha_{\mathrm{A}} < \alpha_{\mathrm{B}}$ if they share the same $k$ parametrisation. The observed relationship between edge tolerance and strain is intuitive because a network that is more heavily loaded is inherently less robust as it is less able to tolerate changes in flow over its edges. As previously discussed, as $\alpha$ tends to infinity, the strain of the network decreases to a minimum, while the strain will be a maximum when $\alpha = 1$. In the example shown in Fig.~\ref{tikz:toy_example_1}, the spring constant is $100 \leq k \leq 1100$. This creates a maximum possible strain of $\varepsilon = 1.86\times 10^{-1}$ and a minimum of $\varepsilon = 3.25\times 10^{-2}$ when the network is on the point of cascade or in a purely topological analysis, respectively.

This simple example provides a theoretical explanation of how strain and tension are related to robustness. Although these are very simple networks, it is evident that even for a complex network, if all lines are loaded to $\alpha = 20$, this will be more robust than the same topology with a loading of $\alpha = 2$, and the strain will be smaller in the first case. There is thus a clear analogy between the physical embedding representation of strain and tension and the `strain' that a power-grid experiences in the form of its edge loading. The hypothesis of the paper is that the giant component will collapse more quickly in networks that have high mean strain or mean tension values when compared to networks with low mean strain or tension, even when the network is not proportionally loaded. 

Appendix.~B provides a more detailed example of SETSe as a robustness metric and demonstrates the strengths of SETSe and Line Load in comparison to system tolerance.

\subsection{Power-grid datasets}
\label{sec:data}
To show the relationships between the different methods and the collapse point of the giant component, this paper uses data from seven power networks: the IEEE power-flow test cases 14, 30, 57, 118, and 300 \cite{noauthor_ieee_nodate}, a network representing the high-voltage UK power-grid, and a synthetic network representing the Texas power-grid \cite{birchfield_grid_2017}. The power-flow test cases are standard networks used for power-flow analysis. The UK power-grid dataset is proprietary but based on the Electrical Ten Year Statement dataset \cite{noauthor_electricity_nodate}. 

Summary statistics of these networks are shown in Table~\ref{tab:net_stats}, in which the columns describe the network type, the number of nodes and edges, the mean betweenness of the network (Betw), the average degree of the nodes (Deg), the assortativity of the network (Assort), the mean clustering coefficient (Clust), the mean node distance (Dist), the number of generator nodes (Gens), and the number of load nodes (Loads). There are some clear patterns in the networks, such as whilst distance, number of generators and number of loads is proportional to network size, Betweenness and Clustering are inversely proportional. The networks are visualised in Appendix.~C.

\begin{table}[ht]
\centering
\begin{tabular}{rlrrrrrrrrr}
  \hline
 & Graph & Nodes & Edges & Betw & Deg & Assort & Clust & Dist & Gens & Loads \\ 
  \hline
1 & IEEE 14 &  14 &  20 & 0.11 & 2.86 & -0.07 & 0.33 & 2.37 &   2 &  11 \\ 
  2 & IEEE 30 &  30 &  41 & 0.08 & 2.73 & -0.09 & 0.18 & 3.31 &   2 &  21 \\ 
  3 & IEEE 57 &  57 &  78 & 0.07 & 2.74 & 0.24 & 0.16 & 4.95 &   4 &  42 \\ 
  4 & IEEE 118 & 118 & 179 & 0.05 & 3.03 & -0.15 & 0.14 & 6.31 &  19 &  99 \\ 
  5 & IEEE 300 & 300 & 409 & 0.03 & 2.73 & -0.22 & 0.10 & 9.94 &  57 & 191 \\ 
  6 & UK grid & 523 & 709 & 0.02 & 2.71 & -0.06 & 0.11 & 11.77 & 165 & 287 \\ 
  7 & Texas grid & 2000 & 2668 & 0.01 & 2.67 & -0.22 & 0.00 & 12.98 & 390 & 1125 \\ 
   \hline
\end{tabular}
\caption{Summary statisitics of the seven power networks used in the study} 
\label{tab:net_stats}
\end{table}

Having introduced the datasets the next sections describes the method of how the load profiles will be generated across the test datasets.

\subsection{Creating load profiles}
\label{sec:strain_robust}
The best way to create load profiles is using real load and demand profiles along with the known edge capacities of the network. However, such information is rare in power-grid datasets. Only the UK and Texas networks come with edge capacities/limits and only the synthetic network of Texas has more than one load profile. Given these limitations a simulated approach to create different line loads is needed. The line loads are created using a simplified redistribution method that is designed to cause skew in both the arithmetic mean and the harmonic mean. This redistribution works by treating the excess capacity of each line -- that is, the capacity above the initial power flow in the line -- as a resource to be re-allocated, thus changing the edge capacities of the lines and concentrating the network capacity into fewer lines.

The edge capacity redistribution process is as follows. Load all edges in the power system to one of the carrying capacity values chosen from the set $\alpha \in \left \{1.005, 1.025, 1.1, 1.2, 1.5, 2, 3, 5, 7, 10, 20 \right \}$. These edge capacities are chosen as they cover the normal range of loading's used in network science analysis of power-grid robustness and continue up to the point where the system has almost no cascades and the attacks become a purely topological analysis \cite{bourne_dont_2019}. The fraction $p$ of the edges with the most excess capacity is selected, where the fraction is chosen from the set $p = \left \{ \frac{1}{\mathrm{V}}, 0.1, 0.2, 0.3, 0.4, 0.5 \right \}$ (where $\mathrm{V}$ is the number of nodes in the network). The fraction of excess capacity removed is taken from the set $f \in \left \{ 0.25, 0.5, 0.75, 0.99 \right \}$, 0.99 is used instead of 1 so that no edge is initialised on the point of failure. The removed excess capacity is then distributed across the fraction of the edges with the least excess capacity, where the fraction is taken from the set $q = \left \{ \frac{1}{\mathrm{V}}, 0.1, 0.2, 0.3, 0.4, 0.5 \right \}$.   The process described above is then performed again except that the excess capacity from the edges with least excess capacity is transferred to the edges with the most excess capacity. 

The result of all combinations of the above parameters creates 3456 edge loading profiles across each network. 

Having described how the load profiles will be generated the next section describes how the attacks and resulting cascading failures will be simulated.

\subsection{Simulating attacks and cascades}

This study considers the robustness of the network not in relation to random failure, but instead through random attack until the collapse of the giant component. Attacking the grid until failure is unlike the random failures that occur in blackouts on the power-grid \cite{bbc_south_2019, noauthor_final_2004,guo_critical_2017, noauthor_final_2004-1} and is more like a targeted attack \cite{glenn_cyber_2016, national_research_council_terrorism_2012, lee_analysis_2016, cherepanov_win32/industroyer_2017}. However, in network science, using the attack style on a network is a well-established method for testing its resilience and robustness \cite{pepyne_topology_2007, arianos_power_2009,buldyrev_catastrophic_2010, kinney_modeling_2005, motter_cascade-based_2002, ouyang_comparisons_2014}, and this paper takes such an approach.

The robustness of a network is a latent property and cannot be directly measured. There are many different approaches to defining metrics for robustness, this paper uses the the mean number of attacks until the loss of the giant component of the network across a number of random attack scenarios. Although, this method has drawbacks, such as not indicating the amount of system power lost, it provides a binary outcome rather then a continuous one. This simplicity lends itself to testing the embedding methods and so is chosen here. To determine whether the giant component has been lost, the Molloy--Reed criterion is used \cite{molloy_critical_1995}. This states that a random network has a giant component when each node has on average more than two connections. For a network with an arbitrary degree distribution, the critical threshold for the existence of the giant component is found at $\left \langle k^2 \right \rangle - 2 \left \langle k \right \rangle >0$, where $\left \langle k^2 \right \rangle$ is the mean-squared degree of the network and $\left \langle k \right \rangle$ is the mean degree of the network. Using the collapse point of the giant component is not without has the disadvantage that it gives no indication of how much of the total power in the system has been lost. However, it is a simple intuitive measure that can be easily measured and compared against the embedding values and thus serves the purpose of testing their applicability for analysing power-grids.

This paper defines the attack simulation using five parameters: physics model, element, attack type, removal method, and load profile. The physics model used in this analysis is DC flow, and the elements attacked are the edges. The attack type is `fixed', meaning that the order in which the edges will be removed is generated before the attack begins; if the next edge in the deletion list has been lost to a cascade, the algorithm continues along the list to the next edge still present in the network. The order in which nodes are removed is sequential. This means that only a single edge is targeted for removal during each round, with any other nodes and edges being lost in the resulting cascade. Cascades occur in the manner described in the introduction, that is if the power flowing over an edge exceeds the capacity of the edge it is removed, and the power flow is recalculated, this process repeats until power flow across edges is within their tolerance and the system is stable.
The code to perform these attacks is included in the R package power-grid Networking \cite{bourne_powergridnetworking_2018}, and all code is available at \url{https://github.com/JonnoB/setse_and_network_robustness}.

Each loading scenario, generated through the process described in section \ref{sec:strain_robust}, is randomly attacked to the point of collapse of the giant component 100 times (effectively a 100 instance Monte-Carlo simulation on each profile) and the total number of attacks averaged, making 345\,600 attack scenarios per network and just over 2.4 million attack scenarios across all seven networks.

Having described how the attacks will be simulated, the next section describes how the appropriateness of the metrics as a proxy for system robustness will be evaluated against the mean collapse point of the network.

\subsection{Assessing ability to predict the collapse point of the giant component}

For each of the networks, the 3456 mean time to giant-component collapse points are predicted using a generalised additive model (GAM) \cite{wood_smoothing_2016, hastie_generalized_1986}. Where the dependent variable is the mean time to collapse and the independent variable being the robustness measure. A GAM has been chosen over other techniques as it is a simple method of non-linear regression that will work even if the forms of the robustness measures are very different.

The robustness measures will be compared using two standard evaluation methods: $R^2$ and the symmetric mean absolute percentage error (SMAPE) \cite{flores_pragmatic_1986}. The coefficient of determination $R^2$ has been chosen because if there is no information about the relationship between the metric and the time to collapse, then knowing that an increase in the metric is strongly correlated with an increase in the collapse time is useful. As the number of attacks to the point of collapse varies greatly depending on the network capacity, it is important that the robustness proxy is accurate relative to the value of the true outcome. This means that whilst large errors are permissible for large values, small errors are necessary for small values, and SMAPE captures this requirement by looking at the error as a percentage, and it will thus be used here. 

The $R^2$ value is defined by $R^2 = 1- \frac{ \mathrm{SS}_{\mathrm{res}} }{  \mathrm{SS}_{\mathrm{tot}} }$, where $\mathrm{SS}_{\mathrm{tot}} = \sum_{i=1}^n (r_i - \bar{r})^2$, and $\mathrm{SS}_{\mathrm{res}} = \sum_{i=1}^n (r_i - p_i)^2$. In the $R^2$ equations $n$ is the total number of observations, $r_i$ is the round at which the giant component collapsed, $\bar{r}$ is the mean round of collapse across all simulations, $\mathrm{SS}_{\mathrm{tot}}$ is the sum of squares between the round of collapse and the mean round of collapse, the value predicted by the model is $p_i$, and the sum of residual squares between the round of collapse and the predicted round is $ SS_{\mathrm{res}}$. The value of the SMAPE is defined as $\mathrm{SMAPE} = \frac{100}{n}\sum_{i=1}^n\frac{\left | p_i -r_i \right |}{(\left | r_i \right | + \left | p_i \right |)/2}$.

Ten repeats of ten-fold cross-validation will be used to see how well the different measures generalise to unseen data. This will mean that 100 models will be created for each network-measure pair, the mean performance, in terms of $R^2$ and SMAPE, across all 100 models will provide the final score for each of the robustness measures and corresponding network. As an example this means there will be 100 GAM's produced for IEEE-118 where elevation is the dependent variable, these 100 models will be evaluated using both $R^2$ and SMAPE as described above.

\section{Time series analysis}
The synthetic Texas grid dataset also includes a time series of hourly data across one year of operation. The method creating this synthetic time series is described in Ref.~\cite{birchfield_grid_2017}. The time series is composed of 8784 periods, for each period the line load and SETSe embedding will be found and 100 attacks to failure performed and the mean attacks to the loss of the giant component recorded. This will result in an additional 878,400 attacks to failure on top of the 2.4 million for the synthetic line load. The resulting data will then be compared using Pearson's correlation \cite{pearson_x._1900}. 

\section{Geospatial analysis}

After finding the embedded node elevations on the base load profile of the UK power-grid and the associated line strain, the relative node elevations will be re-projected back into geographical space, using their coordinates and interpolating between points using kriging \cite{chiles_fifty_2018} with a spherical distance model. This will then be plotted and interpreted to see what qualitative information can be gained from the two methods. Interpolation is not necessary for interpretation, but it makes it easier when looking at geography; an alternative is the use of Voronoi tessellation. 

Having described the method used in this paper the next section provides the results of the experiments described so far.

\section{Results}
\label{sect:results}
The results are broken into three sections. Sec~\ref{ref:robustness_results} explores how well the different methods function as proxies for robustness using seven power-grids across 3456 synthetic loading profiles. Sec~\ref{sec:timeseries} compares the performance of SETSe and Line Load under synthesised normal operation conditions across an entire year on the Texas grid. Finally, in Sec~\ref{sect:UKpowergrid}, SETSe and line loading are compared in geographical space and an interpretation is given.

\subsection{Ability to predict the collapse point of the giant component}
\label{ref:robustness_results}

The loss of the giant component is not a guarantee that the power-grid has lost the majority of its system power. However, across all attack simulations a mean of 79\% of total power was lost with a standard deviation of 9\%, a minimum of 52\% and a maximum of 96\%, indicting that for all distributions the power-grids suffered substantial damage from the attacks, before the giant component collapsed. As would be expected cascade size changed dramatically; for example IEEE-118 had a minimum cascade size of 0.7\% of edges per attack, when loaded with a carrying capacity of 20 and a maximum mean cascade size of 80\% of edges per attack when it had a carrying capacity of 1.005.

\begin{figure}
    \centering
    \includegraphics{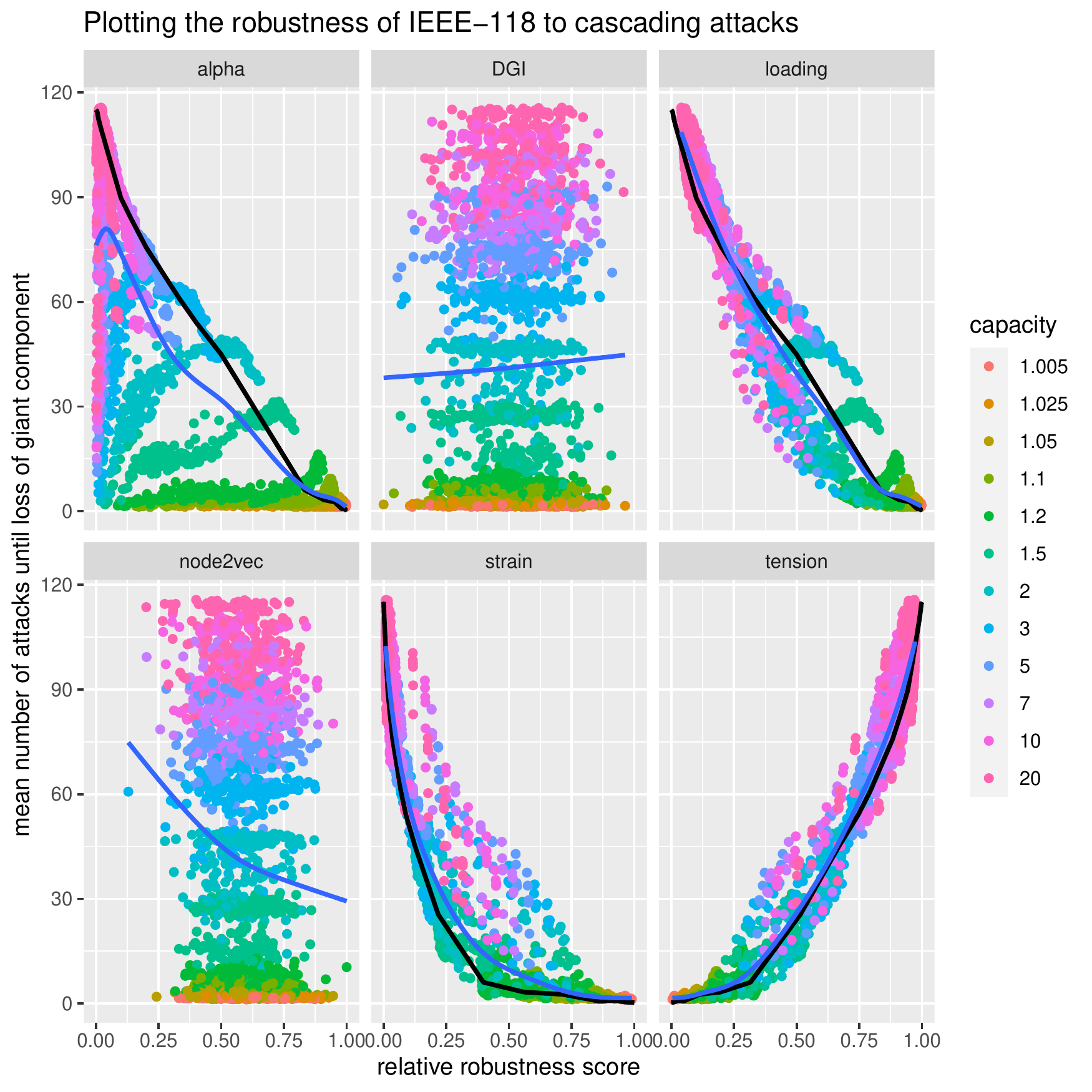}
    \caption{The robustness measures behave very differently when compared to the mean collapse point of the giant component }
    \label{NodesAttacked_Strain_alpha_small_ec_118}
\end{figure}

Fig.~\ref{NodesAttacked_Strain_alpha_small_ec_118} uses IEEE-118 to show the relationship between the number of attacks until the collapse of the giant components and the value produced by the robustness measures; system tolerance, loading, strain, tension, node2vec, and DGI. Elevation has not been included, as it is similar to strain and tension but with a poorer fit. The black lines in Fig.~ \ref{NodesAttacked_Strain_alpha_small_ec_118} show the results when the system is proportionally loaded. The $y$-axis represents the mean point the giant component was lost across 100 simulations, the $x$-axis is the relative robustness score for that measure as discussed in Sec.~ \ref{sec:diff_robust_measures}. Each point in the plot represents a load profile, the colour of the points represent the network carrying capacity. The black line marks the position of proportionally loaded networks (not included for DGI and node2vec due to poor performance).  Although only IEEE-118 is shown the result for all networks are similar.

Mean edge capacity is a poor proxy for robustness when the system is not proportionally loaded. The node2vec and DGI panels of Fig.~\ref{NodesAttacked_Strain_alpha_small_ec_118} show the mean Euclidean distance between adjacent nodes. Both the mean node embeddings in 32-dimensional vector space and the Euclidean distance produce very poor results for node2vec and DGI.

It is interesting to note that $\alpha$ has the properties of Simpson's paradox \cite{simpson_interpretation_1951}, as the correlation across all tolerance values is negative, but it is positive for individual tolerance levels. 

\begin{figure}
    \centering
    \includegraphics{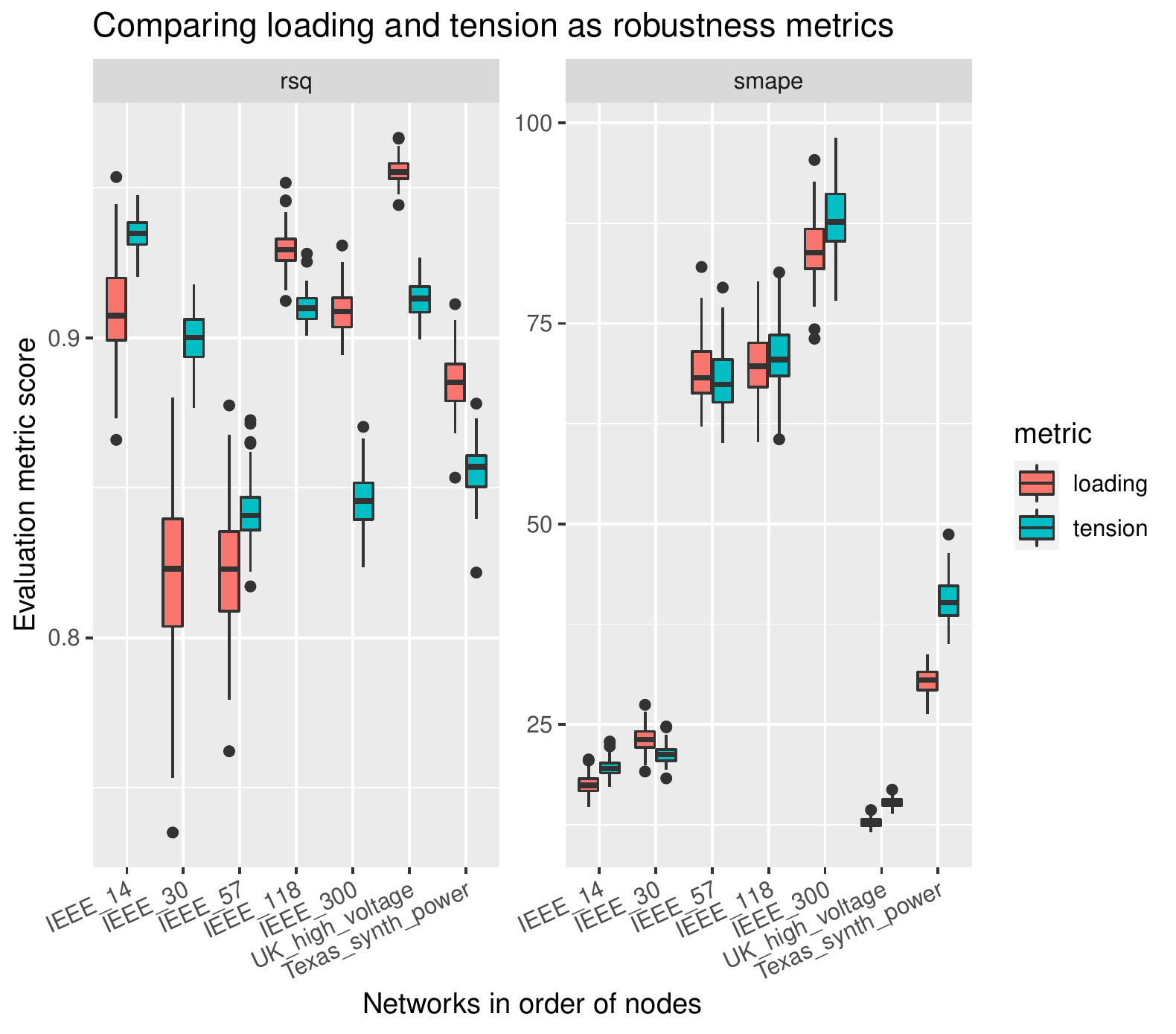}
    \caption{Performance of loading vs tension across 100 cross-validated models for each graph type. Relative performance depends on the network.}
    \label{fig:mod_perf}
\end{figure}

To compare the relative performance of the methods, 100 GAM models were created using ten repeats of ten-fold cross-validation. The resultant evaluation metrics $R^2$ and SMAPE across all 100 models for each graph are shown in Fig.~\ref{fig:mod_perf}. Only tension and loading are shown, as they outperformed the other methods so comprehensively that including all measures would make the figure difficult to interpret. The methods had almost identical scores averaged across the networks, but with larger difference within networks. Both tension and line load have an $R^2$ of 0.89 whilst line load has a slightly lower SMAPE of 43.8 compared to tension's 46.3. Tension was a substantially better predictor of the collapse of the giant component for the smaller networks whilst line load performed better on the larger ones. As a reference, $\alpha$ had an $R^2$ value of 0.5 and a SMAPE of 53, whilst neither node2vec nor DGI could produce models with meaningful predictive power.

A second set of 100 models was made using the same method; however, in this case, the training data used was 20 levels of proportionally loaded networks. These models were only slightly outperformed by the models trained on 3100 examples.

\subsection{Analysing time series data}
\label{sec:timeseries}

Analysis of the timeseries data shows that the system has a mean edge loading of 0.2 a minimum of 0.14 and a maximum of 0.3. This shows that during normal operation the system is lightly loaded relative to it's full capacity although there bottle necks at several points in the network with edges close to tripping.

\begin{figure}
    \centering
    \includegraphics{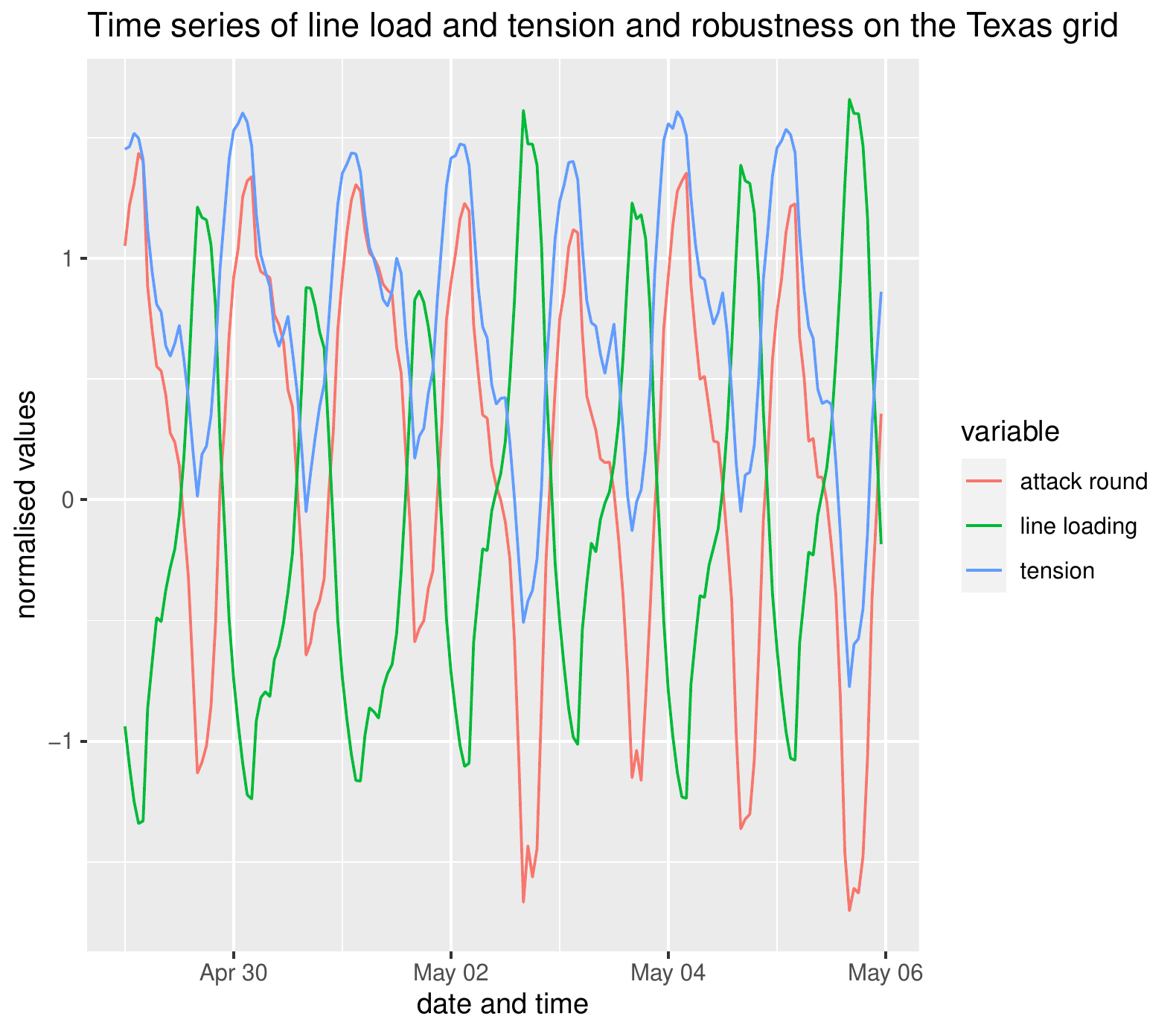}
    \caption{A time series comparing tension and line load with the collapse point of the giant component. Tension and line load mirror system robustness very well, with line load having an almost perfect correlation.}
    \label{fig:timeseries_week}
\end{figure}

Fig.~\ref{fig:timeseries_week} shows a single week of data from the Texas grid dataset. As can be seen both line load and tension have high levels of correlation with the mean point of collapse of the giant component. However, although line load and tension had relatively similar performance on the synthetic data which provided a wide range of loading parameters, using the time series data line load has an almost perfect correlation of $R=0.99$ which is substantially better performance than tension with a correlation of $R=0.80$. The main reason for this is that the normal envelope of operation for the power-grid is within an area of the grid loading parameter space which has lower error than tension. A comparison of the time series and the load profiles used in Sec.~\ref{ref:robustness_results}, can be seen in Appendix.~D.

\subsection{SETSe in geographic space}
\label{sect:UKpowergrid}

\begin{figure}
    \centering
    \includegraphics{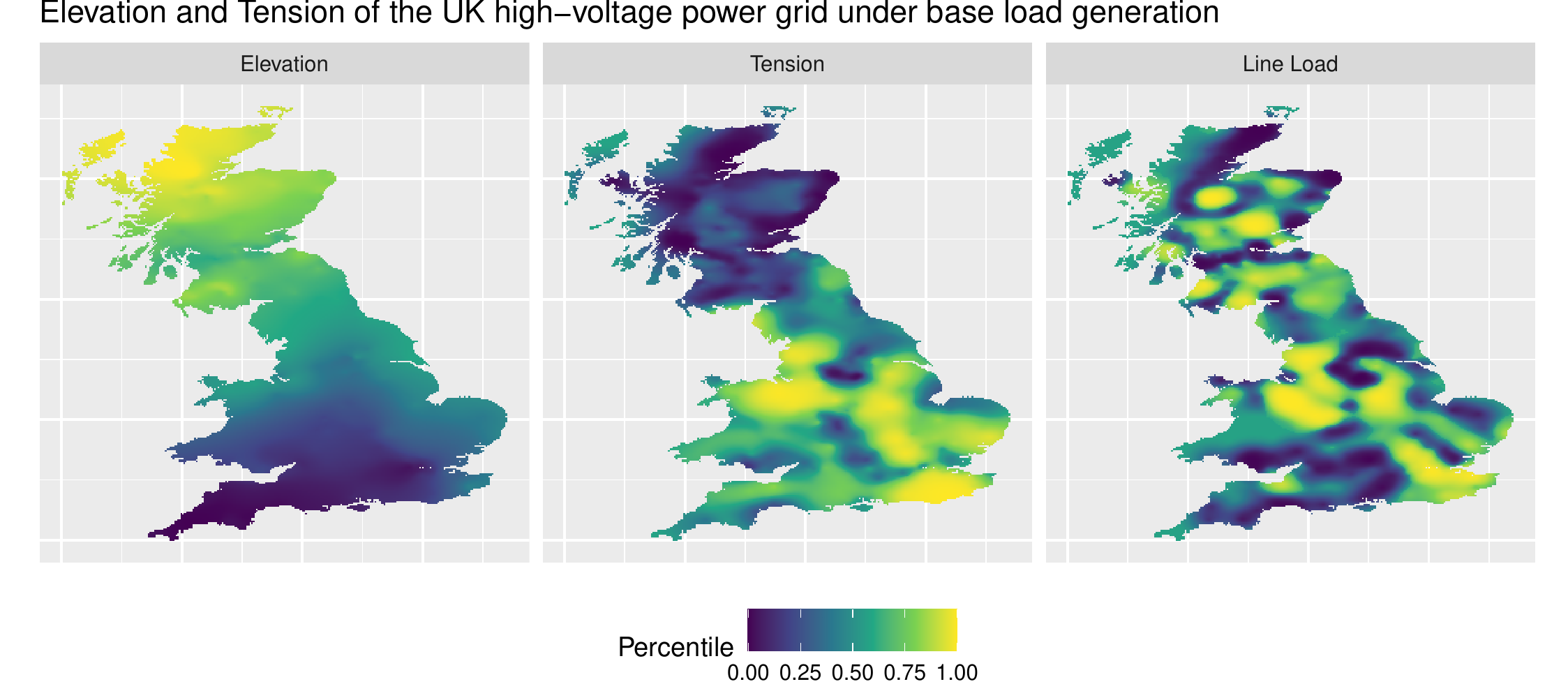}
    \caption{Topology of Great Britain, where elevation is the embedded elevation of the nodes as well as the strain and tolerance of the edges. Values at geographical points between nodes have been interpolated using kriging with a spherical distance model.}
    \label{fig:Brit_kriged_topologly}
\end{figure}

The results of embedding SETSe and line load onto geographical space are shown in Fig.~\ref{fig:Brit_kriged_topologly}. Two notable results visible from the plots are that the embedded elevation tilts from north to south and that there are substantial local differences between the tension and the line load.

The tilt from north to south shows the direction of power flow in the UK mainland. This tilt is caused by the low population density but high level of generation in Scotland, which exports power to the densely populated south-east (England). Within Scotland, we can also see the `valley' created by the high-density Edinburgh--Glasgow urban corridor, which is sandwiched between two high-production, low-demand areas of the Scottish Highlands and south-west Scotland. In the south-east of Britain, the pressure caused by the power demands of London creates a sharp reduction in elevation. This reduction is shown in the left panel of Fig.~\ref{fig:Brit_kriged_topologly} as a change from light blue (close to the 50th percentile) to dark blue (close to the lowest percentiles). The change is also reflected in the tension seen in the centre panel of Fig.~\ref{fig:Brit_kriged_topologly}, as very intense yellow as the nearby nuclear power stations of Sizewell to the north-east of London and Dungeness to the south of London transfer their power through a limited number of high-voltage lines. The tension map also shows that the large metropolitan areas in the Midlands and north of England cause localised tension in the network. In the west of the country, just above Wales, the strain is caused by the cities of Liverpool and Manchester, while further east, the tension is caused by Leeds and Sheffield. The lowest point in the network is Indian Queens in the south-west, while the highest point in the country is Loch Luichart in northern Scotland. 

The differences between the tension and line-load plots highlight one of the strengths of SETSe as an analysis tool. The line-load panel shows the value of each line, but the tension value combines information from the local network to provide a broader picture of local grid support. As a result, we can see that although individual lines may be close to overloading, the local network is well supported, making it robust to the failure of any single line. The correlation between the line tolerance $\alpha$ and the system tension is only 0.45. 

\section{Discussion}

Using seven different networks with a range of nodes from 14 to 2000 and a redistribution method designed to create outliers, this paper found that both tension and line load performed well as proxies for robustness against cascading failure using $R^2$ and SMAPE. The failure of $\alpha$ to be an effective robustness measure was due to the ease with which the value was skewed, and this reinforces the finding of Ref.~\cite{bourne_dont_2019} that $\alpha$ is a poor choice of metric for cascading failure research. In contrast to SETSe, DGI and node2vec lack a theoretical motivation for being able to represent robustness; what's more their loss functions are designed to preserve local structure adding to the difficulty of differentiating between topologically identical networks.

Although both line load and SETSe performed well, there were large differences in their relative performance across the seven networks. SETSe outperformed line load on the smaller networks, where there are few generator nodes. This may be because SETSe is better able to detect the inherent fragility of these networks, reflected in the highly strained edges around the generators. With a large number of generators such a distinction is lost and Line Load becomes the dominant method, this is perhaps as the local smoothing of SETSe reflects inaccurately on the global robustness of the network. Line load outperformed tension by a larger margin on the time series data than compared to the synthetic load profiles, this combined with the distribution observed in Fig.~\ref{NodesAttacked_Strain_alpha_small_ec_118}, suggest that this is due to the network being lightly loaded and the normal operating parameters being relatively close to proportional loading. Overall it appears that the accuracy of the prediction of the collapse point of the giant component is dependent on a range of interacting variables, the analysis of which is beyond the scope of this paper.

In Sec.~\ref{sec:diff_robust_measures} SETSe was described as a method of network smoothing whereby node features are locally smoothed and globally weighted. This effect was clearly shown in Fig.~\ref{fig:Brit_kriged_topologly}, where the map of the UK could be interpreted at local level with relative differences in elevation and line strain, and also global level using absolute elevation. This creates a map that can be used for analysis of `what if' scenarios and can be easily grasped by non-experts. However, this visualisation is only appropriate for planar graphs such as the power-grid, road and rail networks, or river networks.

\section{Conclusions}

Creating methods and metrics that predict the robustness of a network to cascading failure, without using a large number of Monte-Carlo simulations, is challenging. This paper has shown that network embedding methods, such as SETSe, can accurately model the robustness of power-grids to cascading failures. However, not all embedding algorithms are appropriate for the task. DGI and node2vec did not have any predictive ability, even with 32 dimensions in the latent vector space, indicating the loss function used in learning is inappropriately designed for this task. SETSe's success at modelling network robustness due to its local/global feature smoothing and clear analogy between strain/tension and the `strain' on a power line. 

Although SETSe showed that it is possible to represent latent graph properties using embedding methods, it did not outperform line load, which was the strongest performer on the larger grids, and especially so on the time series data. Overall line load has almost no computational cost and exceptional performance under certain loading scenarios; however, SETSe is more flexible and can produce an easily interpretable two-dimensional map. Given their similar $R^2$ and SMAPE scores and the difficulty understanding the relationship between measure performance, topology, and load position, making clear statements about relative performance in a given situation is challenging.

The paper uses 7 different power-grids and simulated the collapse of the giant component of the networks just under 3.3 million times. As such the results are quite robust and to the best of our knowledge this is the most in extensive network-science based study into cascading failure on the power grid.

This paper has shown that graph representation algorithms can be used to analyse abstract or latent concepts such as graph robustness to cascading failure attacks. Although currently such use is in its infancy, further work could greatly improve predictive power. Such work should focus on developing a more sophisticated spring function for SETSe. Likewise designing a loss function for a DeepGraph ML algorithm with a focus other than the preservation of local topology could create a strong predictor. In any case it is clear that being able to integrate a holistic view of topology along with information from the features of the nodes is a powerful and flexible method to generate insight into the latent properties of a network

\section*{Funding}
This work was funded by the EPSRC International Doctoral Scholars - IDS grant (EP/N509577/1). The author declares that no outside body impacted the contents of this study.

\printbibliography
\end{document}